\documentclass[preprint,amsmath,amssymb,aps,longbibliography,floatfix]{revtex4-1}
\usepackage{amsmath,amsfonts,amssymb,graphicx}
\usepackage{bm}
\usepackage[caption=false]{subfig}
\usepackage{float}
\usepackage{mathrsfs}
\usepackage{mathtools}
\usepackage{color,xcolor}
\usepackage[american]{babel}
\usepackage{microtype}

\begin{document}

\title{Experimental simulation of the Parity-Time-symmetric dynamics using photonics qubits}
\author{Wei-Chao Gao}
       \affiliation{School of Artificial Intelligence, Beijing Normal University, Beijing 100875, China}
		\affiliation{School of Science and the State Key Laboratory of Information Photonics and Optical Communications, Beijing University of Posts and Telecommunications, Beijing 100876, China}
\author{ Chao Zheng}
	\affiliation{Department of Physics, College of Science, North China University of Technology, Beijing 100144, China}
\author{Lu Liu}
	\affiliation{School of Science and the State Key Laboratory of Information Photonics and Optical Communications, Beijing University of Posts and Telecommunications, Beijing 100876, China}
\author{Tie-Jun Wang}
	\affiliation{School of Science and the State Key Laboratory of Information Photonics and Optical Communications, Beijing University of Posts and Telecommunications, Beijing 100876, China}
	\author{Chuan Wang}
	\affiliation{School of Artificial Intelligence, Beijing Normal University, Beijing 100875, China}
   \affiliation{School of Science and the State Key Laboratory of Information Photonics and Optical Communications, Beijing University of Posts and Telecommunications, Beijing 100876, China}
\begin{abstract}
The concept of parity-time (PT) symmetry originates from the framework of quantum mechanics, where if the Hamiltonian operator satisfies the commutation relation with the parity and time operators, it shows all real eigen-energy spectrum. Recently, PT symmetry was introduced into optics, electronic circuits, acoustics, and so many other classical fields to further study the dynamics of the Hamiltonian and the energy of the system. Focusing on the dynamical evolution of the quantum state under the action of PT symmetric Hamiltonian, here we experimentally demonstrated the general dynamical evolution of a two-level system under the PT symmetric Hamiltonian using single-photon system. By enlarging the system using ancillary qubits and encoding the subsystem under the non-Hermitian Hamiltonian with post-selection, the evolution of the state can be observed with a high fidelity when the successfully parity-time symmetrically evolved subspace is solely considered. Owing to the effectively operation of the dilation method, our work provides a route for further exploiting the exotic properties of PT symmetric Hamiltonian for quantum simulation and quantum information processing.
\end{abstract}

\maketitle

%

\section{Introduction}

The Hermiticity of the operators is considered as one of the fundamental axioms of quantum mechanics, of which  guarantees  the real energy spectra of the quantum system and the unitary evolution with conserved probability \cite{PhysRevLett.80.5243,Shankar1980Principles}. However, a quantum version of non-Hermitian Hamiltonian satisfying parity-time ($\mathcal{PT}$) symmetry \cite{PhysRevLett.98.040403,Wang_2010} (where $\mathcal{P}$ and $\mathcal{T}$ denote the parity and time reversal operators, respectively) can still exhibit the real energy spectrum and the probability conservation conditions by redefining the inner product \cite{Bender10.1063/1.532860,PhysRevLett.89.270401}. For the case considered here, given that a Hamiltonian \emph{H} with $\mathcal{P}$ ($\mathcal{T}$) symmetry which obeys the relation $\mathcal{P}$\emph{H}$\mathcal{P}$=\emph{H}($\mathcal{T}$\emph{H}$\mathcal{T}$ =\emph{H}) and that with $\mathcal{PT}$ symmetry obeys [\emph{H},  $\mathcal{PT}$]=0. Later, the concept of $\mathcal{PT}$ symmetry was first experimentally observed in electrical circuits system \cite{PhysRevA.84.040101,PhysRevA.85.062122}, and then extended to other systems which usually consists of balanced gain and loss, such as the optical waveguides \cite{Ruschhaupt_2005,PhysRevLett.101.080402,PhysRevLett.103.093902,ruter2010observation,feng2011nonreciprocal}, mechanical systems \cite{doi:10.1119/1.4789549}, optical microcavities \cite{peng2014parity,el2018non},  and optical systems with atomic media \cite{PhysRevLett.110.083604,Yanhong201998,PhysRevLett.117.123601}. These systems exhibit the properties of conservative systems. Recently,  the phenomena for growing interest in exploring novel effects on $\mathcal{PT}$ -symmetric classical optical systems is not only refers to the simulation of $\mathcal{PT}$ -symmetric theory itself but also opening the doors for novel photonic applications, striking examples include the exceptional points \cite{EP-1,Ep-2,EP-3}, unidirectional light transport \cite{feng2011nonreciprocal,peng2014parity} and the single-mode lasers \cite{feng2014single,hodaei2014parity}.

However, experimental study on $\mathcal{PT}$ -symmetric physics in quantum regime remains huge challenge. Some non-trivial and undebatable effects show that the most possible approach for realizing the $\mathcal{PT}$ -symmetric Hamiltonian is to utilize an open quantum system. Yet, it is difficult to achieve a controllable $\mathcal{PT}$-symmetric Hamiltonian by controlling the environment \cite{breuer2002theory}. Some progress associated with $\mathcal{PT}$ -symmetric Hamiltonian has been made with this approach in the system of photons \cite{TangJian20168,XiaoLa201998}, ultracold atoms \cite{LiJiaming2019220}, nitrogen-vacancy(NV) centers \cite{wu2019observation}, nuclear spins \cite{zheng2013observation,PhysRevA.99.062122}, and superconducting qubits \cite{naghiloo2019quantum}. However, unambiguously observing the evolution of a quantum state (qubit) under the the $\mathcal{PT}$-symmetric Hamiltonian has not been realized which is critical not only for further applications of $\mathcal{PT}$ -symmetric theory but also providing great insight for the fundamentals of quantum physics \cite{PhysRevLett.98.040403}.

In this study, we designed a linear optical circuit to simulate the $\mathcal{PT}$ -symmetric and experimentally simulated the evolution of the quantum state under the generalize $\mathcal{PT}$ -symmetric Hamiltonian. The dynamic evolution of the qubit could be observed under the $\mathcal{PT}$ -symmetric Hamiltonian over the time effectively by using the conventional quantum gates and post-selection. Our results show that the evolution of high fidelity of the qubit (quantum state) governed by $\mathcal{PT}$ -symmetric system not only reveals the exotic properties of the $\mathcal{PT}$ theory but also provides a route for further exploiting $\mathcal{PT}$ -theory framework to investigate the fundamental problems.

\section{Methods}

We know, in the realm of quantum computation, the operators must be unitary that means the Hamiltonian is Hermitian.  However, we now construct a two-level
$\mathcal{PT}$-symmetric Hamiltonian Eq.(S1) as a subsystem in a Hilbert space with higher dimensions to simulate the non-unitary operator $U_{\mathcal{PT}}$ (detail see Supplementary Material 1 Eq.(S2)) \cite{Zheng-2018}. The system, used to simulate the $\mathcal{PT}$ time evolution in the case of considering real variables  $r \neq \mu =s$ , consists of a work qubit \textbf{\emph{w}} and an ancillary qubit \textbf{\emph{a}}, which implies a two-qubit system. The initial state of the system is prepared in $|\varphi \rangle _{ini.} = |0\rangle_{\textbf{\emph{a}}}|0\rangle_{\textbf{\emph{w}}}$. First, only the ancillary qubit \textbf{\emph{a}}  of the initial state undergoes the unitary
operation and the work qubit \textbf{\emph{w}} remains unchanged which can be described as  $\textbf{\emph{U}}_{1}$:
\begin{eqnarray}
\begin{split}
\textbf{\emph{U}}_{1} &=\left(\begin{array}{cccc}
cos \  \vartheta_{\textbf{\emph{a}}}  & 0 & -sin \ \vartheta_{\textbf{\emph{a}}} & 0  \\
0 & cos \  \vartheta_{\textbf{\emph{a}}}  & 0 &  -sin \  \vartheta_{\textbf{\emph{a}}}  \\
sin \ \vartheta_{\textbf{\emph{a}}} & 0 & cos \  \vartheta_{\textbf{\emph{a}}} & 0 \\
0 & sin \ \vartheta_{\textbf{\emph{a}}} & 0 & cos \  \vartheta_{\textbf{\emph{a}}} \\
\end{array}\right)     \\
&=
\left(\begin{array}{cc}
cos \  \vartheta_{\textbf{\emph{a}}}  &  -sin \ \vartheta_{\textbf{\emph{a}}}   \\
sin \ \vartheta_{\textbf{\emph{a}}} &  cos \  \vartheta_{\textbf{\emph{a}}}
\end{array}\right)
\otimes I
\end{split}
\end{eqnarray}
where
\begin{eqnarray}
cos \  \vartheta_{\textbf{\emph{a}}}&=&\sqrt{\frac{\omega ^{2}cos^{2}(\omega t/2\hbar)+(\mu +s)^{2}sin^{2}(\omega t/2\hbar)}{\omega ^{2}cos(\omega
t/\hbar)+2(\mu +s)^{2}sin^{2}(\omega t/2\hbar)} }\\
sin \  \vartheta_{\textbf{\emph{a}}}&=& \frac{\sqrt{(\mu -s)^{2}+4r^{2}sin^{2}\theta } \  sin(\omega t/2\hbar)}
{\sqrt{ \omega ^{2}cos(\omega t/\hbar)+2(\mu +s)^{2}sin^{2}(\omega t/2\hbar)}}.
\end{eqnarray}

Then,  the ancillary qubit \textbf{\emph{a}} acts as a control qubit and performs the $U_{2}$ operation on the work qubit \textbf{\emph{w}} only when the \textbf{\emph{a}} is $|0\rangle _{\textbf{\emph{a}}}$:

\begin{eqnarray}
\textbf{\emph{C}}(U_{2})=\left(\begin{array}{cccc}
cos \  \vartheta_{\textbf{\emph{w}}_{1}} & i sin \  \vartheta_{\textbf{\emph{w}}_{1}} & 0 & 0 \\
 i sin \  \vartheta_{\textbf{\emph{w}}_{1}}  & cos \  \vartheta_{\textbf{\emph{w}}_{1}} & 0 & 0 \\
0 & 0 & 1 & 0\\
0 & 0 & 0 & 1 \\
\end{array}\right)
=
\left(\begin{array}{cc}
U_{2} &  0 \\
0 & I
\end{array}\right)
\end{eqnarray}
where
\begin{eqnarray}
cos \  \vartheta_{\textbf{\emph{w}}_{1}}&=&\sqrt{\frac{\omega ^{2}}
{\omega ^{2}+(\mu +s)^{2}tan^{2}(\omega t/2\hbar)} }\\
sin \  \vartheta_{\textbf{\emph{w}}_{1}}&=&-\frac{(\mu +s)tan(\omega t/2\hbar)}
{\sqrt{\omega ^{2}+(\mu +s)^{2}tan^{2}(\omega t/2\hbar)} }
\end{eqnarray}

And that performs the $U_{3}$ operation on the work qubit \textbf{\emph{w}} only when the \textbf{\emph{a}} is $|1\rangle _{\textbf{\emph{a}}}$:

\begin{eqnarray}
\textbf{\emph{C}}(U_{3})=\left(\begin{array}{cccc}
1 & 0 & 0 & 0 \\
0 & 1 & 0 & 0 \\
0 & 0 & cos \  \vartheta_{\textbf{\emph{w}}_{2}} & -i sin \  \vartheta_{\textbf{\emph{w}}_{2}} \\
0 & 0 &  i sin \  \vartheta_{\textbf{\emph{w}}_{2}}  & -cos \  \vartheta_{\textbf{\emph{w}}_{2}} \\
\end{array}\right)
=
\left(\begin{array}{cc}
I & 0  \\
0 & U_{3}
\end{array}\right)
\end{eqnarray}
where
\begin{eqnarray}
cos \  \vartheta_{\textbf{\emph{w}}_{2}}&=&\frac{2 r sin \theta}
{\sqrt{(\mu -s)^{2} +4 r^{2} sin^{2}\theta}}\\
sin \  \vartheta_{\textbf{\emph{w}}_{2}}&=&-\frac{\mu -s}
{\sqrt{(\mu -s)^{2} +4 r^{2} sin^{2}\theta}}
\end{eqnarray}

\begin{figure*}[ht!]
	\centering
\centerline{\includegraphics[width=0.7\textwidth]{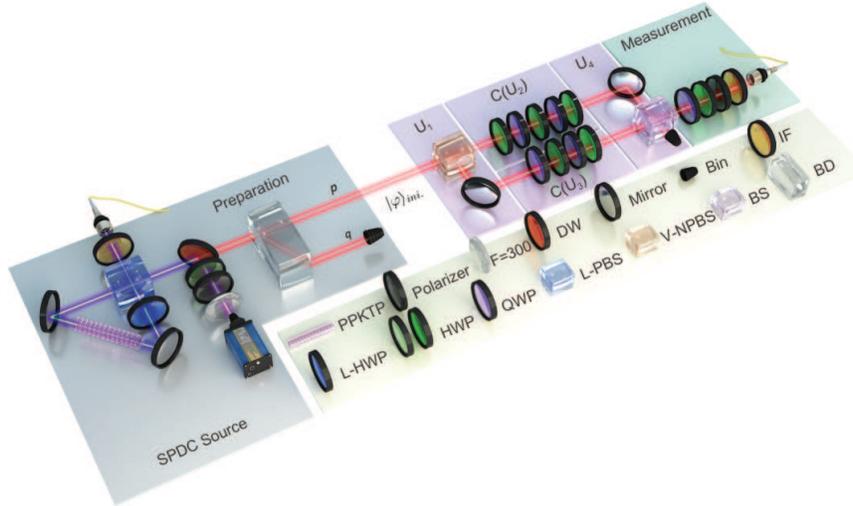}}
\caption{Experimental setup. The polarization of a single photon generated form polarization Sagnac interferometer type-II down-conversion  is first prepared with two degrees of freedom (polarization and path ) via a BD. The polarization state and path state are used to encode as work qubit and ancillary qubit, respectively, which is the initial input state $|\varphi \rangle _{ini.}$. The middle block, which contains a NPBS, two combined-array  half- and quarter- wave plat groups and BS,  implements the $U_{1}$, $ C(U_{2}) $, $ C(U_{3}) $ and $ U_{4} $ operations for two qubits. The labels $p$ and $q$ represent the ancillary path states. The output state is measured by post-selection. Key to components: NPBS, non-polarizing Beamsplitter; BD, beam displacer; L-HWP, long-broadband half-wave plate; L-PBS, long-broadband polarizing beam splitter; DW, dichroic mirror.}
\end{figure*}

At final, a Hadamard operation $U_{4}= \textbf{H} $ is applied to the ancillary qubit \textbf{\emph{a}}. After the above series of unitary operations, the initial state $|\varphi \rangle _{ini.} = |0\rangle_{\textbf{\emph{a}}}|0\rangle_{\textbf{\emph{w}}}$ will evolve into:
\begin{eqnarray}
\begin{split}
|\varphi \rangle _{fin.} = &\frac{c}{\sqrt{2}}[|0\rangle_{\textbf{\emph{a}}} e^{-i \frac{t}{\hbar}H_{\mathcal{PT}}} |0\rangle_{\textbf{\emph{w}}}\\
&+\frac{1}{c} |1\rangle_{\textbf{\emph{a}}}(cos \  \vartheta_{\textbf{\emph{a}}}  U_{2} - sin \  \vartheta_{\textbf{\emph{a}}} U_{3} ) |0\rangle_{\textbf{\emph{w}}}]
\end{split}
\end{eqnarray}
where \textit{c} is a non-zero coefficient
\begin{eqnarray}
c= \frac{\omega e^{i \frac{t}{\hbar} r cos \theta }}
{\sqrt{\omega ^{2}cos(\omega t/\hbar)+2(\mu +s)^{2}sin^{2}(\omega t/2\hbar)} }
\end{eqnarray}
and only affects the possibility to obtain the ancillary qubit \textbf{\emph{a}} being in state $|0\rangle_{\textbf{\emph{a}}}$ but does not influence  the law of $\mathcal{PT}$-symmetric evolution that work qubit \textbf{\emph{w}} obeys.

Until now, the two-qubit system is observed. If the ancillary qubit \textbf{\emph{a}} is measured in state $|0\rangle_{\textbf{\emph{a}}}$, the work qubit \textbf{\emph{w}} does evolve into state $e^{-i \frac{t}{\hbar}H_{\mathcal{PT}}}|0\rangle_{\textbf{\emph{w}}}$, which indicates that the evolution associated with the work qubit \textbf{\emph{w}} is charged by the $\mathcal{PT}$-symmetric Hamiltonian in Eq. (S2).

\section{The experimental simulation of dynamical evolution under Parity-Time-symmetric Hamiltonian }

The sketch the experimental configuration is shown in Fig.1 which includes three modules: the preparation module, the evolution and the detection part. In the preparation module,  we use a 30-mm-long, 2-mm-wide,1-mm-thick flux-grown PPKTP crystal ($\texttt{Raicol Crystals}$) with a grating period of 10.0 $ \mu $m for frequency-degenerate type-$\Pi $ quasi-phase-matched collinear parametric down-conversion, which is pumped by 45-mW vertically polarized beam at 405 nm (\texttt{Omicron}) to generate heralded single photons in the framework of Sagnac interferometer. One of the entangled photons is set as target qubit for coincidence detection while the other photon, with polarization degree of freedom(\emph{H}(horizontal) and  \emph{V}(vertical)), is identified as working qubit \textbf{\emph{w}} for evolution. Then, a beam displacer(BD) is used to generate the spatial mode degree of freedom(marked with path \emph{p} and path \emph{q}) which is encoded as ancillary qubit \textbf{\emph{a}}. The initial state of the two-qubit system is prepared in the form $|\varphi \rangle _{ini.} = |0\rangle_{\textbf{\emph{a}}}|0\rangle_{\textbf{\emph{w}}}$, wherein the \emph{H}(\emph{V}) and path \emph{p}(path \emph{q}) are encoded as $|0\rangle _{\textbf{\emph{w}}}(|1\rangle_{\textbf{\emph{w}}})$ and$|0\rangle _{\textbf{\emph{a}}}(|1\rangle _{\textbf{\emph{a}}})$, respectively. It is worth mentioning that  $|0\rangle _{\textbf{\emph{w}}} (|1\rangle_{\textbf{\emph{w}}})$ and $|0\rangle _{\textbf{\emph{a}}} (|1\rangle _{\textbf{\emph{a}}})$ are synergistically related(work together) due to the constructive contribution of BD.

In the evolution module, four subsequent stages are involved. In the first stage, the unitary operation $U_{1}$ acts on the ancillary qubit \textbf{\emph{a}}(Eq.(1)). That is, the photon in path \emph{p}($|0\rangle_{\textbf{\emph{a}}}$) would be separated into two paths [path \emph{p}($|0\rangle_{\textbf{\emph{a}}}$) and path \emph{q}($|1\rangle_{\textbf{\emph{a}}}$)] according to the T/R ratio of the NPBS, while the other degree of freedom [horizontal component($|0\rangle _{\textbf{\emph{w}}}$)] remains unchanged. The following two steps, the polarization degrees of photons(horizontal component($|0\rangle _{\textbf{\emph{w}}}$) are controlled by path  \emph{p}($|0\rangle_{\textbf{\emph{a}}}$) and path \emph{q}($|1\rangle_{\textbf{\emph{a}}}$) on the respective paths where the corresponding unitary operations $C(U_{2})$(Eq.(4)) and $C(U_{3})$(Eq.(7)) are performed, respectively. In order to guarantee the accurate evolution of $\mathcal{PT}$-symmetric Hamiltonian, the combined-array  half- and quarter- wave plat group is used to implement the operations $U_{2}$ and $U_{3}$ (detail see Supplementary Material 2). In the final steps, a BS followed by the half- and quarter- wave plats group is used which performs the $U_{4}$(Hadamard gate) operation on path \emph{q}($|1\rangle_{\textbf{\emph{a}}}$) and path  \emph{p}($|0\rangle_{\textbf{\emph{a}}}$), wherein a relative phase between these two paths can be ignore and does not change the law of $\mathcal{PT}$-symmetric evolution. In the detection module, post-selection on the path \emph{p} would let the work qubit \textbf{\emph{w}} collapse to the final state $|\varphi \rangle ^{\textbf{\emph{w}}}_{fin.}= e^{-i \frac{t}{\hbar}H_{\mathcal{PT}}} |0\rangle_{\textbf{\emph{w}}} $, which denotes the final state after the $\mathcal{PT}$-symmetric evolution. It is dependent on the ancillary qubit \textbf{\emph{a}} whether is in state $|0\rangle_{\textbf{\emph{a}}}$. Finally, the state information is reconstructed via a quantum state tomography setup, which consists of a quarter- wave plate, a half-wave plate, a polarizer and an interference filter(IF)(detail see Supplementary Material 4).

\begin{figure}[ht!]
	\centering
\begin{minipage}{0.85\linewidth}
\centerline{\includegraphics[width=1\textwidth]{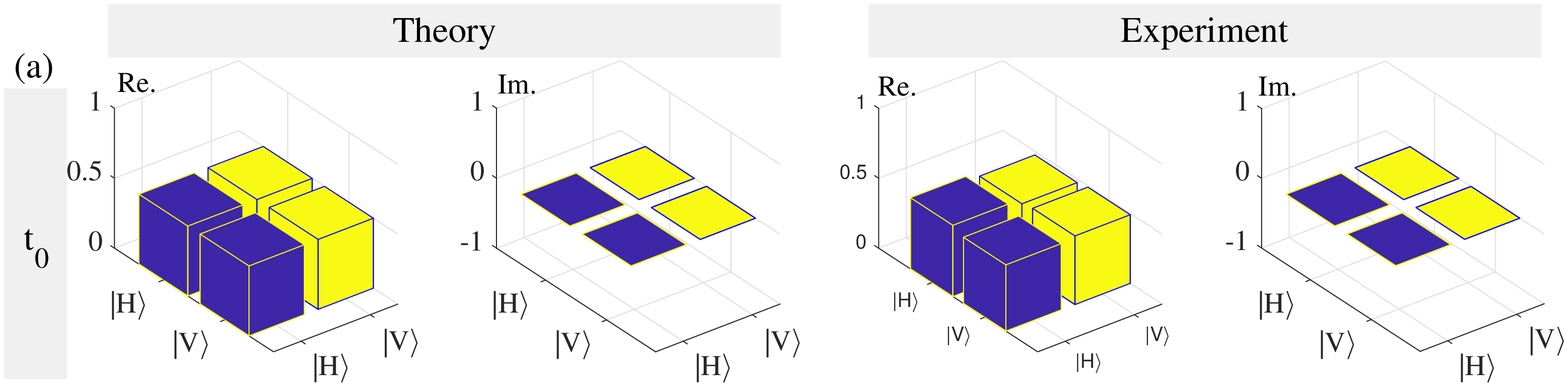}}
\centerline{}
\end{minipage}
\begin{minipage}{0.85\linewidth}
\centerline{\includegraphics[width=1\textwidth]{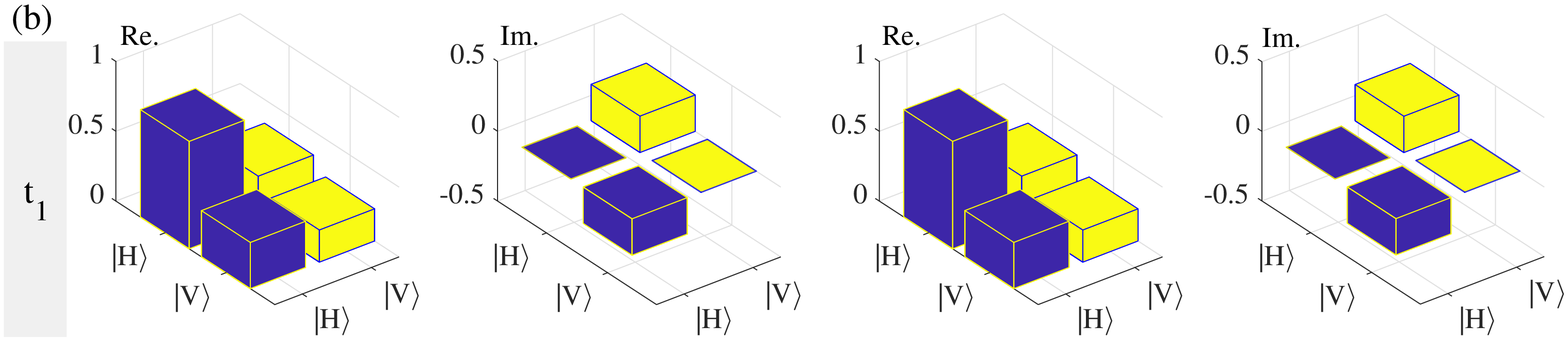}}
\centerline{}
\end{minipage}
\begin{minipage}{0.85\linewidth}
\centerline{\includegraphics[width=1\textwidth]{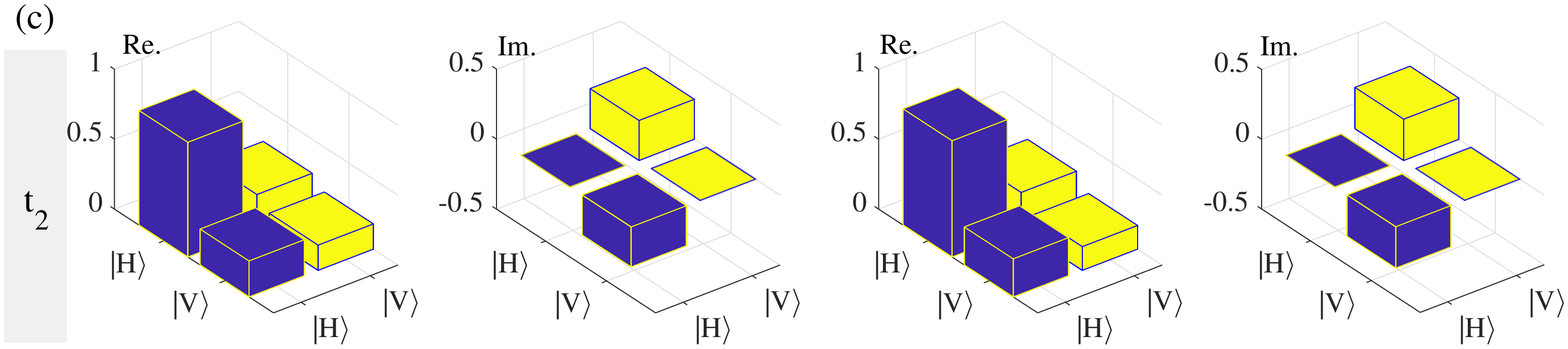}}
\centerline{}
\end{minipage}
\begin{minipage}{0.85\linewidth}
\centerline{\includegraphics[width=1\textwidth]{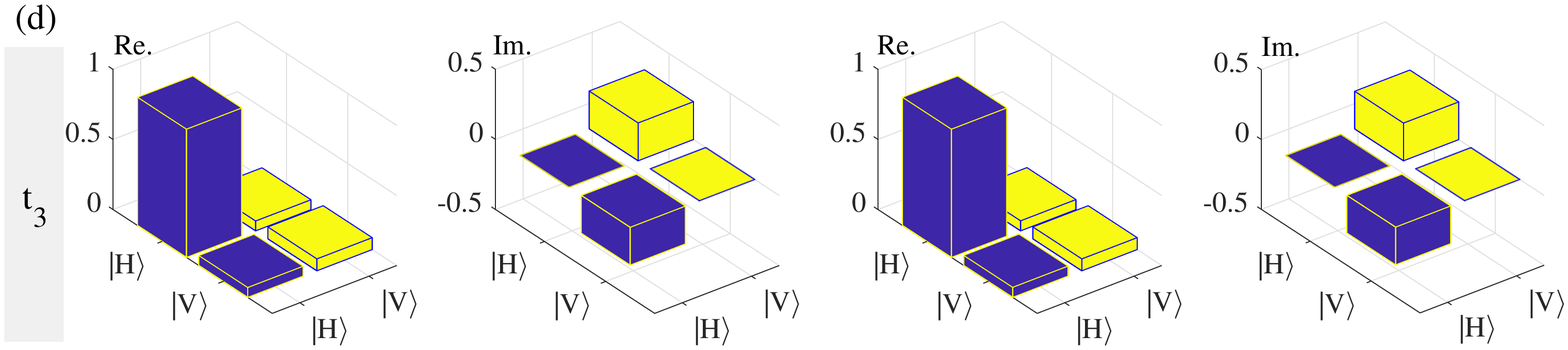}}
\end{minipage}
\caption{The results of the dynamic evolution of state under the $U_{\mathcal{PT}}$ ($r$=2, $ \mu $=$s$=1 and $ \theta $=$ \pi /8 $) with respect to time $t$. (a) -(d) shows the density matrix of the results of output state $|\varphi \rangle ^{\textbf{\emph{w}}}_{fin.} $ at different moments ($t_{0}$, $t_{1}$, $t_{2}$ and $t_{3}$) when the input state is always prepared in $|\varphi \rangle _{ini.}$. In each row (for a certain moment $t_{i}$), the left two columns (the real and imaginary) and the right two columns (the real and imaginary) respectively represent the values calculated theoretically and measured experimentally and they are almost identical. The average difference between these two matrices at each moment is only 0.011 $\pm$ 0.005 (the standard deviation is calculated by the Monte Carlo analysis).}
\end{figure}

\begin{table}[t] \tiny
	\renewcommand\arraystretch{1.2}
	\centering
	\newcommand{\tabincell}[2]{\begin{tabular}{@{}#1@{}}#2\end{tabular}}
	\begin{tabular}[t]{| c | c | c | c |}
	\hline
	\textbf{Time} &  \textbf{U}$_{1}$ & \textbf{U}$_{2}$& \textbf{U}$_{3}$\\
	    \cline{1-4}	
 $t_{0}$ & $  NPBS_{0}(T/R=1:0)$ & \tabincell{c}{ $ HWP^{h}(@0^\circ) \rightarrow QWP^{h}(@0^\circ)\rightarrow $ \\ $  HWP^{v}(@0^\circ)\rightarrow $ \\ $ QWP^{h}(@0^\circ)\rightarrow  HWP^{h}(@0^\circ)$ } & $QWP^{h}(@0^\circ)\rightarrow $ \\
		\cline{1-3}		
$t_{1}$ & $  NPBS_{1}(T/R=4:1)$ & \tabincell{c}{ $ HWP^{h}(@0^\circ) \rightarrow QWP^{h}(@0^\circ)\rightarrow$ \\ $ HWP^{v}(@20.4^\circ)\rightarrow $ \\ $ QWP^{h}(@0^\circ)\rightarrow  HWP^{h}(@0^\circ)$ } & $  HWP^{v}(@0^\circ) \rightarrow $ \\
		\cline{1-3}		
$ t_{2} $ &$  NPBS_{2}(T/R=3:1)$ & \tabincell{c}{ $ HWP^{h}(@0^\circ) \rightarrow QWP^{h}(@0^\circ)\rightarrow $ \\ $ HWP^{v}(@24.5^\circ)\rightarrow $ \\ $ QWP^{h}(@0^\circ)\rightarrow  HWP^{h}(@0^\circ) $  }&$  QWP^{h}(@0^\circ)\rightarrow $ \\
		\cline{1-3}			
$ t_{3} $ &$  NPBS_{3}(T/R=2:1)$ & \tabincell{c}{  $ HWP^{h}(@0^\circ) \rightarrow QWP^{h}(@0^\circ)\rightarrow $ \\ $ HWP^{v}(@33.8^\circ)\rightarrow $  \\ $ QWP^{h}(@0^\circ)\rightarrow  HWP^{h}(@0^\circ)$ } &$  HWP^{h}(@0^\circ)$ \\
		\hline
	\end{tabular}
	\caption{Specific operations of the dynamic evolution of two qubits under $U_{\mathcal{PT}}$. See Method for details of the construction of the $U_{\mathcal{PT}}$.}
\end{table}

During the experiment, dynamical evolution of the qubit under the $\mathcal{PT}$-symmetric Hamiltonian is observed with respect to different time $t$. Here, the $\mathcal{PT}$-symmetric Hamiltonian $U_{\mathcal{PT}}$ in Eq.(S1) is set to be in the form for which $r$=2, $ \mu $=$s$=1 and $ \theta $=$ \pi /8 $. Due to the inevitable experimental imperfections, we present the dynamical evolution of the $\mathcal{PT}$-symmetric Hamiltonian along with four time points: $ t_{0}=0 s $, $ t_{1}=0.7876 s $, $ t_{2}=0.9894 s $, $ t_{3}=1.5521 s $(here $s$=s$ \cdot \hbar$), which corresponds to the operations of $U_{1}$, $C(U_{2})$ that are fabricated with different T/R ratio structures of NPBS and rotating angle of wave plates, experimentally. $C(U_{3})$ only depends on the parameter setting of $U_{\mathcal{PT}}$ and does not change with time. Here Talbe 1 shows the specific experimental operation settings for the dynamical evolution of the $\mathcal{PT}$-symmetric Hamiltonian with respect to time $t$.

The comparison between the theoretical results and the experimental results for the evolved subspace is shown in Fig.2, which corresponds to the Hamiltonian $U_{\mathcal{PT}}$ with different values of $t$ . From Fig. 2(a) to Fig. 2(d), the dynamic evolution process of the initial work qubit \textbf{\emph{w}} $|\rho \rangle ^{\textbf{\emph{w}}}_{ini.}$ are shown which undergoes the $\mathcal{PT}$-symmetric Hamiltonian $U_{\mathcal{PT}}$ over time $t_{0}$, $t_{1}$, $t_{2}$, $t_{3}$, where $|\rho \rangle ^{\textbf{\emph{w}}}_{ini.}$ can be obtained by tracing out the ancillary qubit \textbf{\emph{a}} from $|\varphi \rangle _{ini.}$. The real and imaginary parts of the final reconstructed density matrix via a quantum state tomography are found to be the same as those of the theoretically predicted density matrix, which confirms the perfect combination of our theoretical framework and experimental results. Subsequently,  the formula $F(\rho _{the.}, \rho _{exp.})=Tr(\rho _{the.} \rho _{exp.})/ ( \sqrt{Tr(\rho _{the.}^{2})} \sqrt{Tr(\rho _{exp.}^{2})})$ \cite{Jozsa} are employed to calculate the fidelities between the theoretical expectation values and the experimental results with respected to different $t$ as shown in Fig.3 (a), from which we can calculate that the average value of  fidelity is  0.9992 $\pm$ 0.0002.

Furthermore, the dynamical evolution of the state under $U_{\mathcal{PT}}$ are also studied via monitoring $P_{|0\rangle_{\textbf{\emph{w}}}}$ in the time range from $t_{0}$ to $t_{3}$ which is shown in Fig.3 (b). Considering the form of $U_{\mathcal{PT}}$ in Eq.(S1), the dynamics of $P_{|0\rangle_{\textbf{\emph{w}}}}$  could be always observed in the range of unbroken $\mathcal{PT}$ symmetry. It can be obtained that the measured results (blue line) dependence of the state evolution under $U_{\mathcal{PT}}$ with $t$ are well coincide with the theoretical predictions (red line), and the system eventually approaches to a steady state.

\begin{figure}[ht!]
	\centering
  \begin{minipage}[b]{0.45\textwidth}
\includegraphics[width=\textwidth]{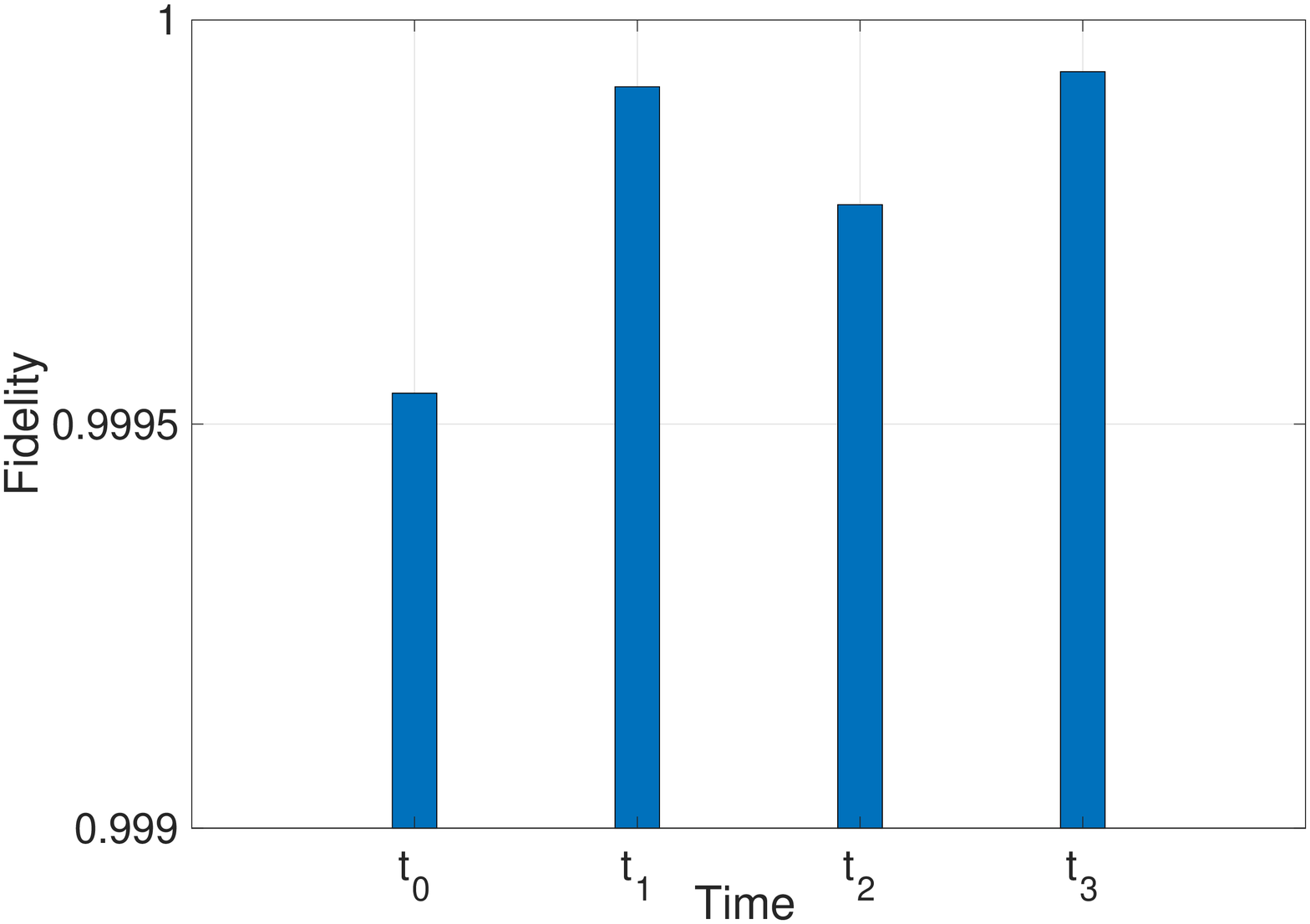}
\end{minipage}
\begin{minipage}[b]{0.45\textwidth}
\includegraphics[width=\textwidth]{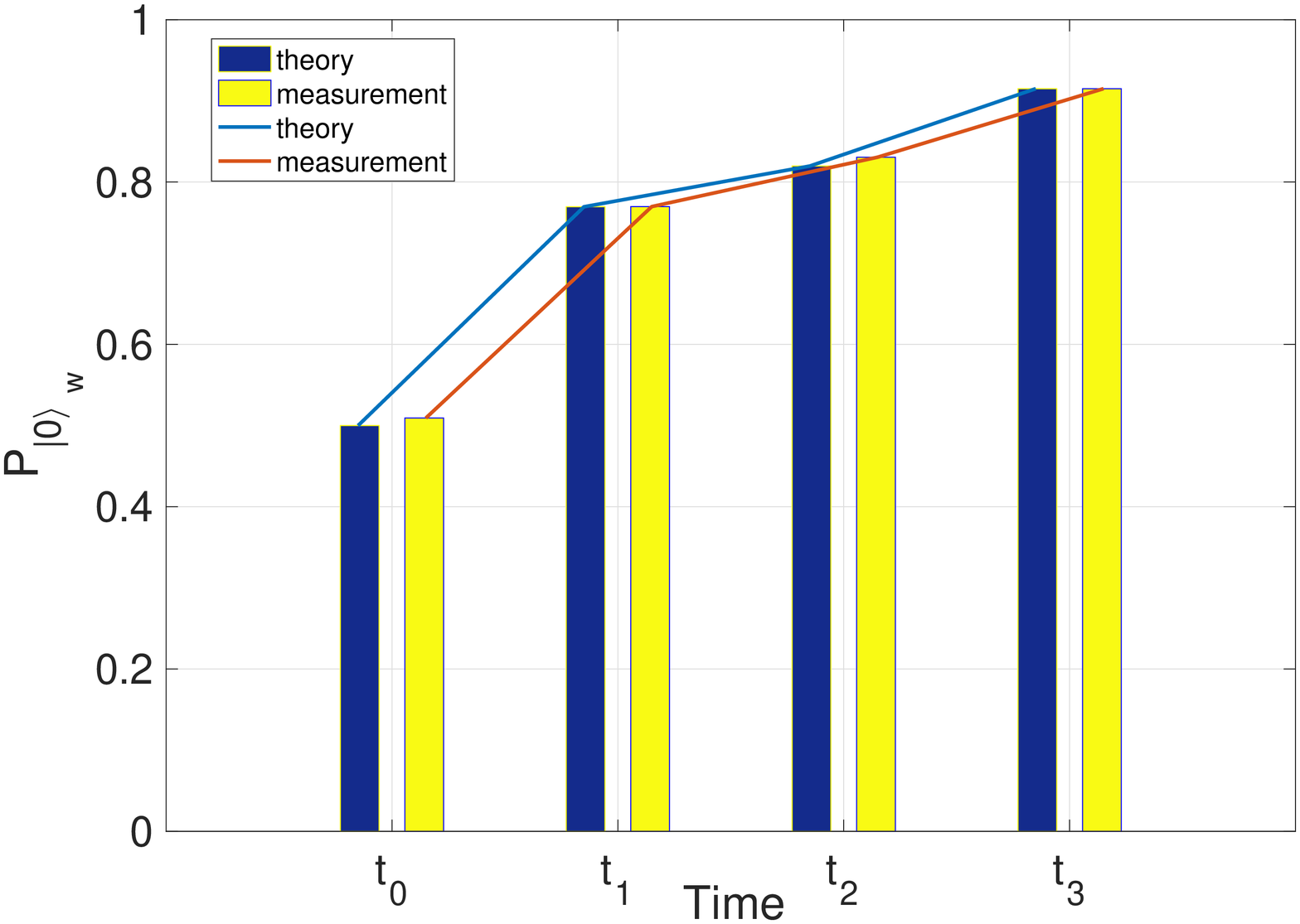}
\end{minipage}
\caption{(a) The fidelities between the theoretical expectation and experimental results with respected to different $t$. (b) Probability($P_{|0\rangle_{\textbf{\emph{w}}}}$) of the state of work qubit($|0\rangle_{\textbf{\emph{w}}}$), as a function of time $t_{i}$, are obtained under the
guarantee of $U_{\mathcal{PT}}$.}
\end{figure}

\section{Summary}

Considering the practical condition in our experiment, the successfully evolved subspace is only a simulation of the $\mathcal{PT}$ -symmetric system as a part of the full Hermitian system on account of: 1) Observing the evolution of the broken -$\mathcal{PT}$ -symmetry zone is poor at our optical system; 2) Most quantum computation and quantum simulation problems can be implemented in the unbroken -symmetry zone, such as no-signalling principle \cite{TangJian20168}. In fact, we do not have to construct a complete $\mathcal{PT}$ Hermitian, because our purpose is to observe and characterize the evolution of single-photon qubit through the framework of $\mathcal{PT}$, which is expected to be used to implement one of the basic systems of dedicated quantum information processing.

In summary, we experimentally investigate the quantum simulation of the dynamical behaviors under the $\mathcal{PT}$ -symmetric Hamiltonian using a single-photon system.  The $U_{\mathcal{PT}}$ operator is effectively simulated in a subsystem of a full Hermitian system using post-selection. The results show that the  state during the evolution can be observed with a high fidelity when the $\mathcal{PT}$ -symmetrically evolved subspace is solely considered. Owing to the effectively operation of the dilation method, our work provides a route for further exploiting the exotic properties of parity-time symmetric Hamiltonian for quantum simulating and quantum computing.

\section*{Acknowledgment}

 This work is supported by the Ministry of Science and Technology of the People's Republic of China (MOST) (2016YFA0301304); the National Natural Science Foundation of China through Grants Nos. 61622103, 11705004, 61701035. and 61671083; the Fundamental Research Funds for the Central Universities(BNU)  and the Fok Ying-Tong Education Foundation for Young Teachers in the Higher Education Institutions of China (Grant No. 151063).

\bibliographystyle{apsrev4-1}
\bibliography{ptref}

\end{document}